\documentclass[11pt,letterpaper]{article}

\usepackage{amsmath,amssymb,amsthm}
\usepackage{deepthink}

\usepackage[nameinlink]{cleveref}
\usepackage[
  backend=biber,
  natbib=true,
  style=alphabetic,
  maxbibnames=100,
  minbibnames=100,
  maxcitenames=2,
  mincitenames=2
]{biblatex}
\usepackage[utf8]{inputenc} 
\usepackage[T1]{fontenc}    
\usepackage{hyperref}       
\usepackage{url}            
\usepackage{booktabs}       
\usepackage{amsfonts}       
\usepackage{nicefrac}       
\usepackage{microtype}      
\usepackage{xcolor}         
\usepackage{bm}
\usepackage{multirow}
\usepackage{tcolorbox}
\usepackage{amsmath}
\usepackage{cleveref}
\usepackage{graphicx}
\usepackage{xcolor}
\usepackage{wrapfig}

\title{See What I See, Know What I Think: Dense Latent Communication Across Heterogeneous Agents}

\authorblock{
  \href{https://chicychen.github.io/}{\textbf{Siyi Chen}}\textsuperscript{1,2}, 
  \href{https://xiaoyanzhang1.github.io/}{\textbf{Xiaoyan Zhang}\textsuperscript{1}}, 
  \href{https://scholar.google.com/citations?user=kUqmflsAAAAJ&hl=zh-TW}{\textbf{Meng Wu}\textsuperscript{1}}, 
  \href{https://research.nvidia.com/person/jonathan-tremblay}{\textbf{Jonathan Tremblay}\textsuperscript{2}}, 
  \href{https://research.nvidia.com/person/valts-blukis}{\textbf{Valts Blukis}\textsuperscript{2}}, 
  \href{https://sbirchfield.github.io/}{\textbf{Stan Birchfield}\textsuperscript{2}}, \\
  \href{http://vision.jhu.edu/rvidal.html}{\textbf{Rene Vidal}\textsuperscript{3}},
  \href{https://www.colorado.edu/cs/alvaro-velasquez}{\textbf{Alvaro Velasquez}\textsuperscript{4}},
  \href{https://lsjxjtu.github.io/}{\textbf{Sijia Liu}\textsuperscript{5}}, 
  \href{https://qingqu.engin.umich.edu/}{\textbf{Qing Qu}}\textsuperscript{1}
}

\affiliation{
  \textsuperscript{1}University of Michigan \, $\cdot$ \quad \textsuperscript{2}NVIDIA \, $\cdot$ \, 
  \textsuperscript{3}University of Pennsylvania
  $\cdot$ \, 
  \textsuperscript{4}University of Colorado Boulder
  $\cdot$ \, 
  \textsuperscript{5}Michigan State University
}

\abstracttext{
Language-model agents build internal representations of the information they observe and the reasoning they perform. Sharing these representations offers a way to communicate both source information and reasoning across agents. For agents built from different models, this requires aligning their representations while preserving information useful to the receiver. We study this problem through KV-cache communication, examining how an agent uses internal states shared by other agents, with or without direct access to the information that other agents observed. A controlled self-communication study shows that cache pruning causes substantially greater degradation when the receiving agent lacks access to that information. We use this finding to guide dense cross-model cache alignment, combining positional disentanglement and KV-group transformations with reconstruction followed by generation training. Across six directed Qwen3 pairs, aligned caches improve in-domain accuracy over text communication when both agents observe the same input, with fewer estimated inference FLOPs. Experiments with three-agent document sharing and Mistral-to-Qwen transfer further demonstrate that aligned caches can carry information across both multiple separate observations and different model families.
}


\date{\today}
\correspondence{\href{mailto:siyich@umich.edu}{siyich@umich.edu}}
\resources{\href{https://chicychen.github.io/dense-hetero-latent-mas/}{\textbf{Project page}} 
}

\headerlogo{}{}

\begin{document}

\makeDeepthinkHeader
\vspace{-.2in}
\begin{figure}[h!]
    \centering
    \includegraphics[width=1\textwidth]{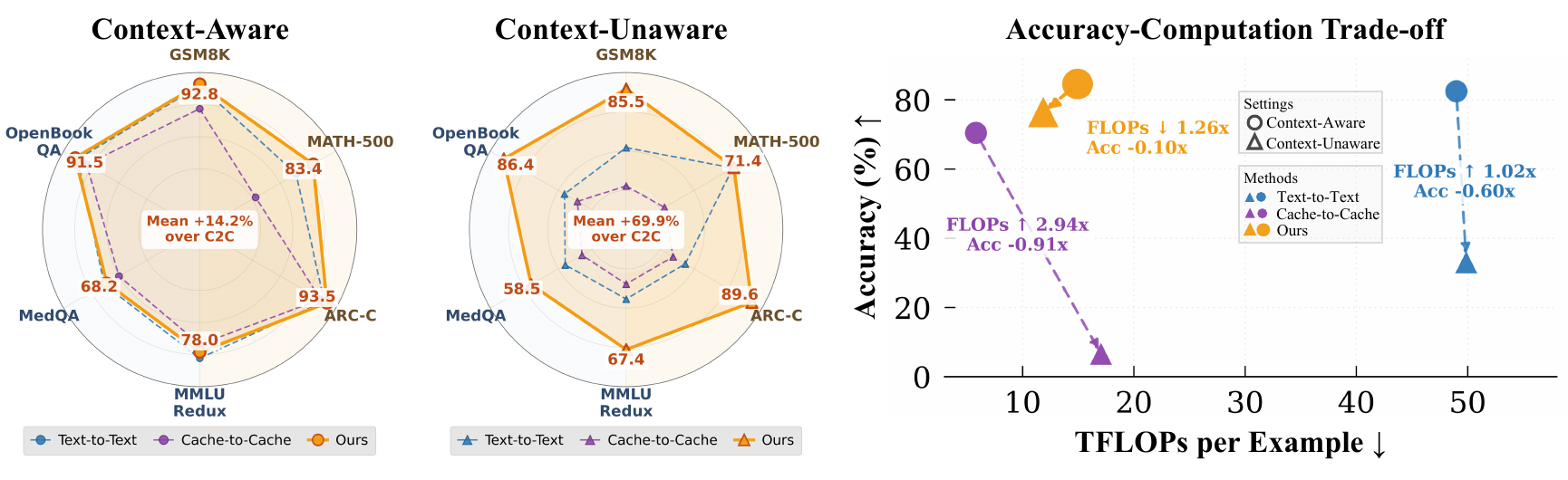}
    \vspace{-.3in}
    \caption{\textbf{Dense latent communication across heterogeneous agents.} Aligned sender KV caches support receiver generation with or without direct access to the source input. We evaluate task accuracy and estimated inference FLOPs.}
    \label{fig:teaser1}
\end{figure}

\newpage
\tableofcontents

\newpage

\section{Introduction}

\begin{wrapfigure}{R}{0.55\textwidth}
    \centering
    \includegraphics[width=\linewidth]{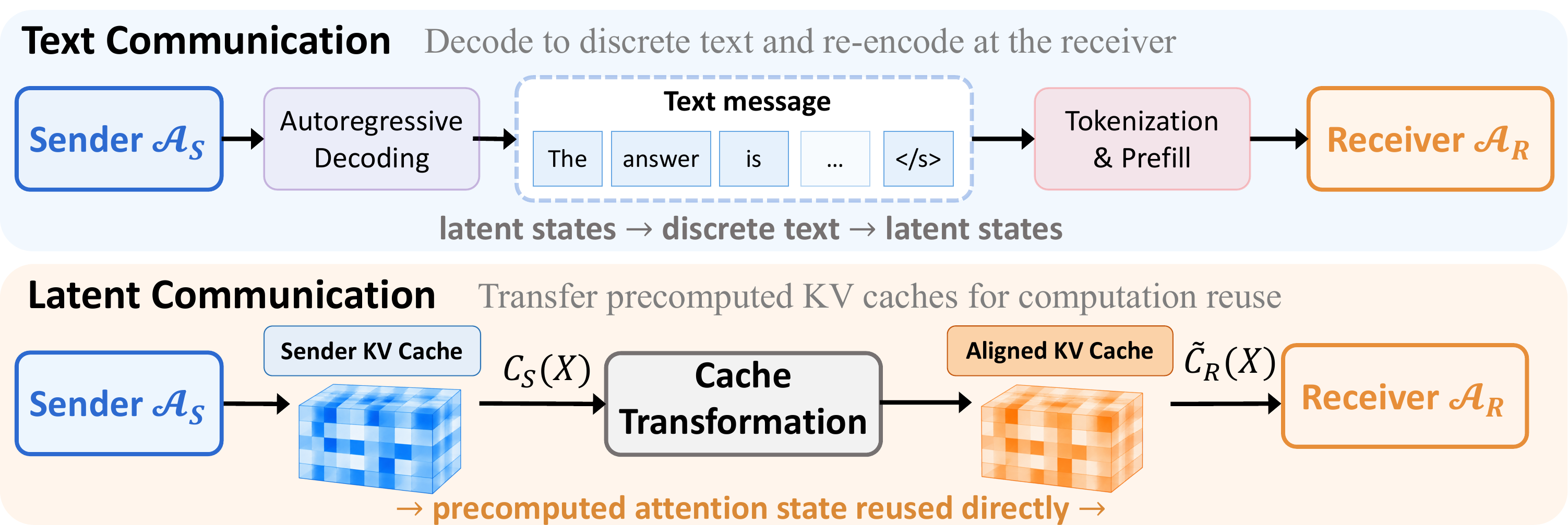}
    \caption{Text communication generates and re-encodes messages; latent communication transfers aligned KV caches.}
    \label{fig:text_latent}
\end{wrapfigure}

Language-model agents can contribute both what they observe and what they infer: a retriever may supply evidence, while a planner develops intermediate conclusions~\citep{tran2025multi,wang2026llm}. Their internal representations offer a channel for sharing source information and the computations built on it~\citep{zheng2025thought,du2025enabling,zou2025latent}. Text provides a flexible, model-agnostic interface, but expressing these contributions as messages requires generation and may omit useful details. Latent communication (\Cref{fig:text_latent}) instead aims to make the sender's internal states directly usable by another agent.

KV caches retain contextual states computed by the sender and provide an interface that the receiver's attention layers can use directly. Work on cache sharing, selective transmission, and learned projection provides ways to reuse these states across agents~\citep{shi2025kvcomm,fu2025cache,li2026less,liu2024droidspeak}. For example, Cache-to-Cache learns projection and fusion between models that process a shared input~\citep{fu2025cache}. We study how to make these states useful across models, including when the receiver has not seen the sender's input. This requires handling differences in cache dimensions, layer organization, positional conventions, and learned representations~\citep{fu2025cache,heo2026cross}, while preserving task-relevant content. To examine how transfer depends on the receiver's available context, we compare context-aware communication, where the receiver sees the source input, with context-unaware communication, where it relies on the transferred message. We also consider agents with complementary private evidence (\Cref{subsec:communication-regimes}). Appendix~\ref{app:related_work} discusses related interfaces in more detail.

Our starting point is a simple question: how much of a cache must be retained when the receiver has not seen the source input? In a controlled Qwen3-4B self-communication study, context-aware performance can be largely preserved with a small subset of selected KV groups, whereas context-unaware performance requires substantially denser retention. This difference motivates an alignment objective that preserves the contents of the cache, alongside an objective for downstream generation. Our approach combines positional disentanglement, per-KV-group transformations and gates, and two-phase training: receiver-cache reconstruction followed by generation supervision.

Across six directed pairs of Qwen3 models, the learned interface supports communication with and without receiver access to the source input. In the context-aware setting, it improves accuracy over text communication on all three in-domain benchmarks across all six pairs, while using fewer estimated inference FLOPs. In the context-unaware setting, it outperforms both text communication and Cache-to-Cache in 33 of the 36 Qwen3 task--pair combinations in our evaluation. In the three-agent HotpotQA task, the receiver combines two independently encoded document memories, improving answer exact match and F1 over both text communication and a receiver given all documents directly in all three evaluated configurations. Mistral-to-Qwen transfer also improves over text communication on the three in-domain benchmarks in both context-aware and context-unaware settings. Together, these results demonstrate that aligned caches can support collaboration across model sizes, model families, and distributed observations. Our main contributions are summarized as follows.
\begin{itemize}[leftmargin=*]
    \item \textbf{Observation-dependent analysis:} a controlled KV-group retention study comparing communication with and without receiver access to the source input.
    \item \textbf{Cross-model cache alignment:} a dense KV-cache interface trained to reconstruct receiver caches and support downstream generation.
    \item \textbf{Empirical evaluation:} transfer across six directed Qwen3 pairs and six benchmarks, comparing task accuracy and inference FLOPs, with additional tests of distributed document memories and cross-family transfer.
\end{itemize}

\section{Communication Setup}
\label{sec:background}

\subsection{KV-Cache Interface}
\label{subsec:MAS-KV-Cache}
We consider language-model agents whose local states reflect both their observations and model-specific computations. Each directed interaction transfers a locally constructed cache into a receiving agent's representation space; a receiver may consume several independently constructed memories. We first define the interface from a sender language model $\mathcal A_S$ to a receiver $\mathcal A_R$. Given source input $\bm X_S$, the sender constructs a KV cache
\begin{equation}
\mathcal C_S(\bm X_S)
=\{(\bm K_S^{(l,g)}(\bm X_S),\bm V_S^{(l,g)}(\bm X_S))\}_{l,g},
\end{equation}

where $l$ indexes layers and $g$ indexes KV groups. A learned transformation produces a receiver-compatible cache $\widetilde{\mathcal C}_R=\mathcal T_{\bm\theta}(\mathcal C_S(\bm X_S))$. The receiver generates an output using this cache and its locally available input $\bm X_R$:
\begin{equation}
p_{\mathcal A_R}(\bm y\mid\bm X_R,\widetilde{\mathcal C}_R).
\end{equation}

Copies of the same model can share compatible cache representations directly, whereas different models generally require alignment of cache structure and learned representations. We review standard attention operations, tensor dimensions, and KV-cache construction in Appendix~\ref{app:attention_preliminaries}.

\subsection{Communication Regimes}
\label{subsec:communication-regimes}
We distinguish three regimes according to receiver access to task information:
\begin{itemize}[leftmargin=*]
    \item \textbf{Context-aware:} both agents observe the source input, $\bm X_S=\bm X_R=\bm X$. The transferred cache supplies the sender's representation of an input also available to the receiver.
    \item \textbf{Context-unaware:} only the sender observes $\bm X$; $\bm X_R=\emptyset$ denotes the absence of source input, rather than the absence of a generation prompt. The message is the receiver's only access to source-side task information, including the question and any answer options.
    \item \textbf{Partial-aware:} agents have complementary local observations, and communication conveys evidence unavailable to the receiver. Question visibility depends on the task: in our three-agent evaluation (\Cref{sec:three_agent}), two senders encode private supporting documents before the question arrives, while the receiver later observes the question and its own distractor documents.
\end{itemize}
The aware and unaware regimes provide controlled comparisons of source-input availability. The partial-aware setting evaluates how a receiver uses evidence held by other agents, including memories constructed before its question arrives.

For $m$ incoming memories, the receiver conditions on
\begin{equation}
p_{\mathcal A_R}\!\left(\bm y\mid\bm X_R,
\operatorname{Merge}(\widetilde{\mathcal C}_{1\to R},\ldots,\widetilde{\mathcal C}_{m\to R})\right).
\end{equation}

Here each cache is transformed independently. Our evaluated extension uses two senders and one receiver; Merge concatenates transformed content caches along the sequence dimension and applies receiver positional encoding over the combined sequence (\Cref{sec:method}).

\begin{figure}[t]
    \centering
    \includegraphics[width=0.88\linewidth]{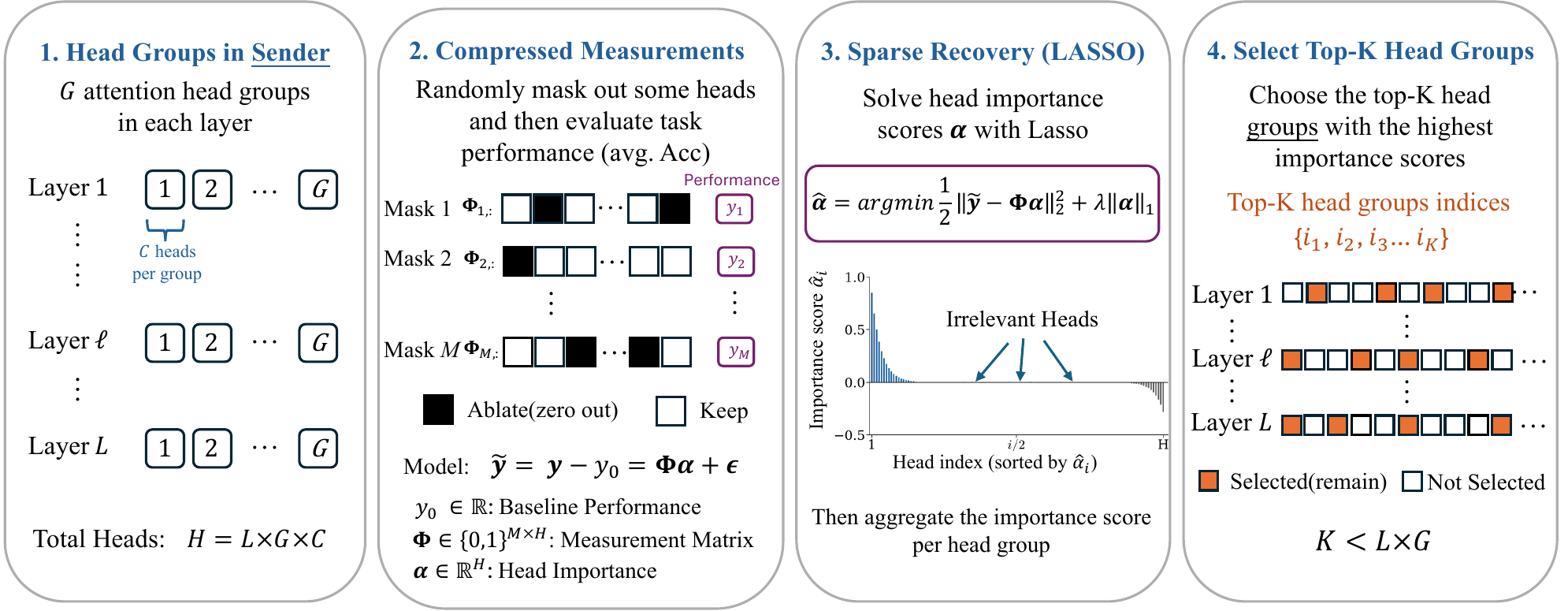}

    \caption{Post-hoc ranking and retention procedure. Random sender query-head ablations supply performance measurements for a Lasso surrogate. Aggregated coefficients rank KV groups for a separate sweep that zeros non-selected cache groups.}
    \label{fig:compressed_sensing}
\end{figure}

\section{How Receiver Context Changes Cache Requirements}
\label{sec:analysis}

\paragraph{Controlled setup.}
We ask whether receiver access to the source input changes how much of a transmitted KV cache is needed for successful communication. To isolate this question from cross-model alignment, we use Qwen3-4B as both sender and receiver on GSM8K, MATH-algebra, and ARC-Challenge. The sender encodes the source input without autoregressive generation. We compare the context-aware and context-unaware regimes defined in \Cref{subsec:communication-regimes}. This self-communication study probes cache-retention sensitivity; heterogeneous alignment is evaluated separately.

\subsection{Retention Experiment}
\label{sec:cs_analysis}
\Cref{fig:compressed_sensing} outlines our adaptation of the compressed-sensing head-ablation method extended from \citet{bair2026compressed}. Let $\Phi\in\{0,1\}^{M\times N}$ encode $M$ random ablations over the sender's $N$ query heads across all layers: $\Phi_{mi}=1$ means head $i$ is ablated in measurement $m$. Given the resulting task accuracies $\mathbf y$ and unablated accuracy $y_0$, we fit a Lasso surrogate to the centered measurements $\tilde{\mathbf y}=\mathbf y-y_0\mathbf 1$:
\begin{equation}
(\hat{\mathbf x},\hat b)
=\arg\min_{\mathbf x,b}
\frac{1}{2M}\|\tilde{\mathbf y}-\Phi\mathbf x-b\mathbf 1\|_2^2
+\alpha\|\mathbf x\|_1,
\label{eq:cs_ranking}
\end{equation}

where $b$ is an intercept and $\alpha$ controls sparsity. More-negative coefficients indicate heads whose ablation predicts a larger accuracy decrease. For KV group $g$ in layer $l$, we aggregate its associated query-head coefficients as
\begin{equation}
s_{l,g}=\sum_{h\in\mathrm{group}(g)}\hat x_{l,h},
\label{eq:cs_group_score}
\end{equation}

where $\hat x_{l,h}$ indexes the fitted vector by layer and head. We retain the $K$ groups with the smallest scores out of $288$ and compare against random group selection under both receiver regimes. The additive surrogate ranks groups by estimated ablation sensitivity. Non-selected key and value entries are zeroed without changing tensor shapes, so the sweep measures retention sensitivity rather than transmitted-byte savings. Appendix~\ref{app:cs_setup} gives the ablation implementation, fitting settings, and evaluation details.

\begin{figure}[t]
    \centering
    \includegraphics[width=0.88\linewidth]{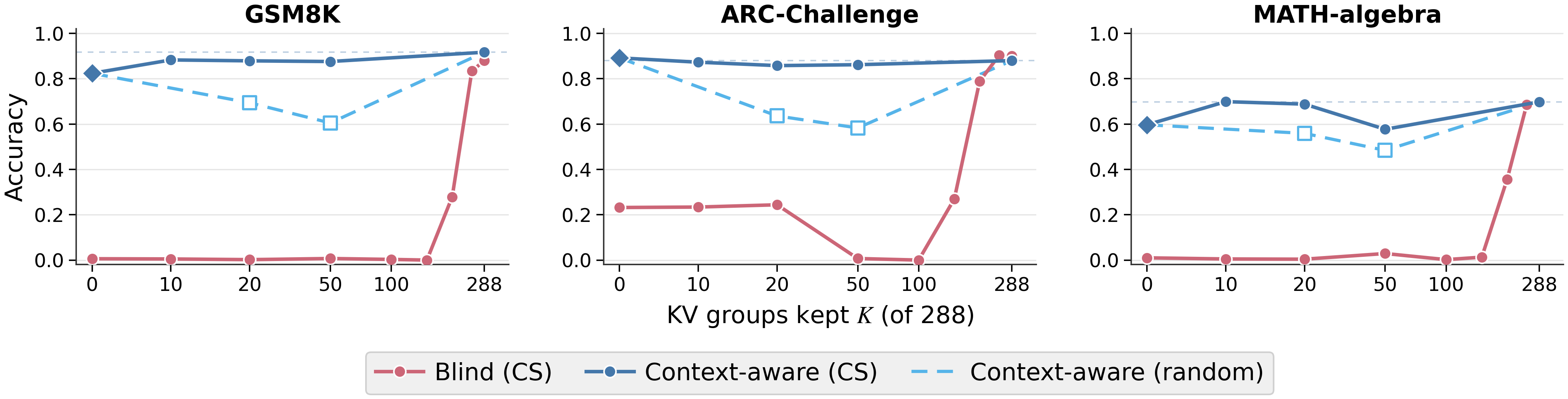}

    \caption{\textbf{Receiver context changes cache-retention sensitivity.} Qwen3-4B self-communication accuracy versus retained KV groups ($K$ of $288$). Solid blue: context-aware CS selection; open blue squares: context-aware random selection; solid red: context-unaware CS selection. The context-aware $K=0$ point is the receiver-only baseline. Non-selected cache groups are zeroed.}
    \label{fig:sparse-dense-contrast}

\end{figure}

\subsection{Findings and Implications}
\label{subsec:observation}
\Cref{fig:sparse-dense-contrast} distinguishes supplementing an available input from conveying missing source information. With access to the source input, the receiver maintains high accuracy with only $10$ of $288$ KV groups: $88.3\%$, $87.3\%$, and $69.9\%$ on GSM8K, ARC-Challenge, and MATH-algebra, versus $91.7\%$, $88.0\%$, and $69.8\%$ with all groups. Without that input, accuracy remains low through $K=150$ and recovers at higher retention. The contrast shows that access to the source input reduces dependence on the transmitted cache. When that input is unavailable, successful generation depends on retaining a broader set of cache groups. Receiver-only references use separate evaluation runs, so we base this interpretation on retention sensitivity within each regime (Appendix~\ref{app:cs_setup}).

Performance also varies nonlinearly with retention, and adding groups does not consistently help in the context-aware regime. Group selection matters: at $K=50$, selected groups achieve $87.6\%$ on GSM8K, compared with $60.5\%$ for random selection, although the two perform similarly at $K=10$. These patterns suggest unequal, context-dependent group contributions and possible interactions. Dense coverage need not imply uniform weighting.

\begin{figure}[t]
    \centering
    \includegraphics[width=.8\linewidth]{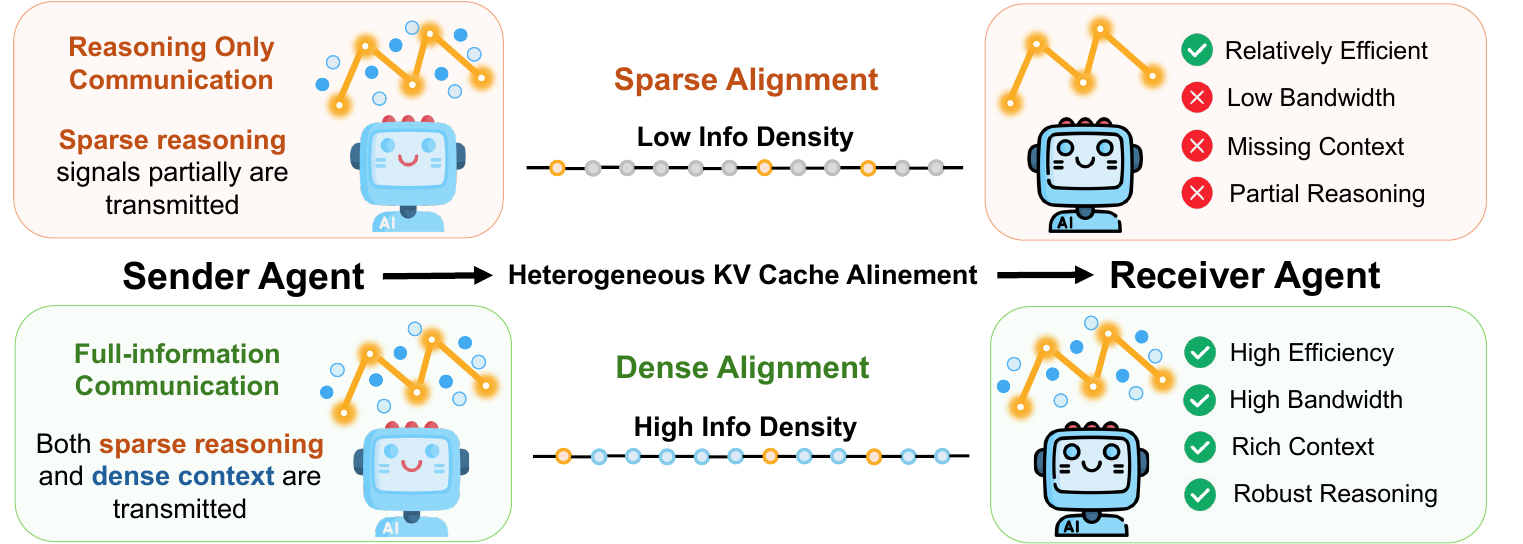}
    \caption{Conceptual contrast between selective cache transfer and dense alignment. Our retention study measures sensitivity to KV-group pruning under different receiver observation regimes; it motivates preserving broad source information when the receiver lacks the source input.}
    \label{fig:sparse-dense}
\end{figure}

The contrast in \Cref{fig:sparse-dense} motivates three complementary choices in \Cref{sec:method}. \emph{Dense cache reconstruction} encourages preservation of broad source content. \emph{Separate KV-group transformations and learnable gates} accommodate differences in group relevance within that coverage. \emph{Generation supervision} refines these transformations and weights for the receiver's task, beyond matching cache values alone. This combines broad alignment with fine-grained, task-directed refinement.

\section{Design of Dense Latent Communication}\label{sec:method}

\begin{figure}[t]
    \centering
    \includegraphics[width=0.92\linewidth]{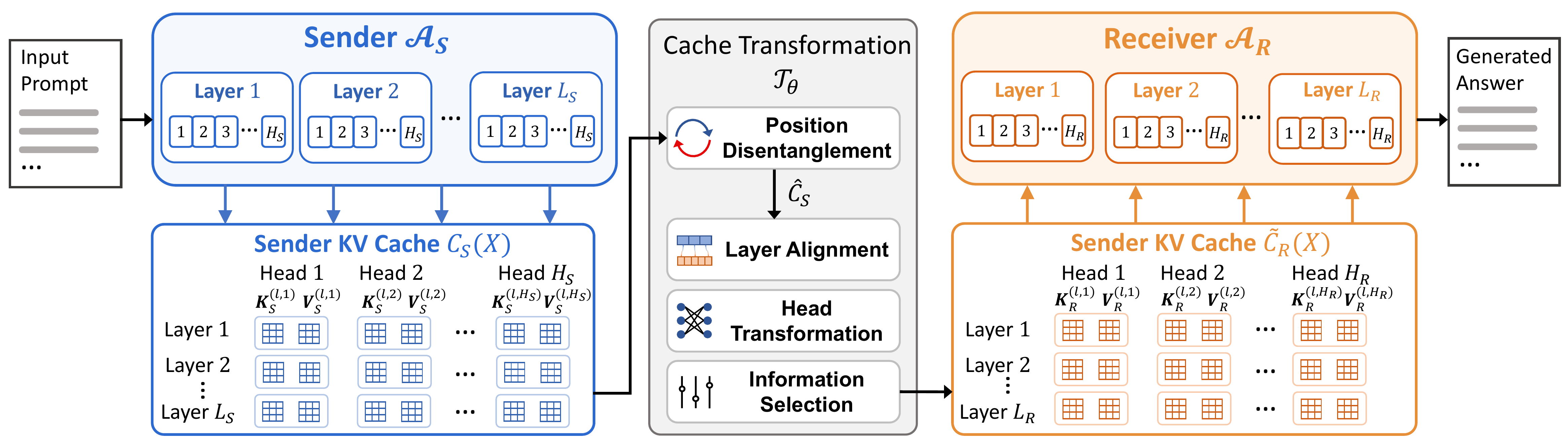}

    \caption{Overview of dense alignment: transform sender KV caches into receiver-compatible caches with positional disentanglement, per-KV-group transformations, and two-phase training.}
    \label{fig:dense_latent}
\end{figure}

The retention study in \Cref{sec:analysis} motivates broad cache coverage with group-specific adaptation: missing source information calls for denser retention, while group selection and receiver context affect its usefulness. We use \emph{dense alignment} to mean learning receiver-compatible key and value states across receiver layers and KV groups, with cache reconstruction as well as downstream generation supervision. Learned gates weight individual groups within this interface; they do not impose a fixed sparse transmission budget. We learn a directed cache transformation $\mathcal T_{\bm\theta}$ from a sender's representation space to a receiver's, using positional disentanglement, layer alignment, KV-group transformations, and gates (\Cref{fig:dense_latent}). The same interface supports a single sender or several independently encoded memories. We first describe training for the two-agent benchmarks, then explain the extension to two senders.

\subsection{Two-Phase Training}

Generation loss supervises the receiver's output but does not directly constrain how closely the transferred cache represents the source input. We therefore first train the adapter to reconstruct the receiver's cache, then tune it for downstream generation. This design gives the adapter both a representation target and a task objective.

\paragraph{Phase I: receiver-cache reconstruction.}
We run both agents on the same source context $\bm X$ and obtain the sender and receiver caches
\begin{equation}
\mathcal{C}_m(\bm X)
=
\left\{
\left(
\bm K_m^{(l,g)}(\bm X),
\bm V_m^{(l,g)}(\bm X)
\right)
\right\}_{l,g},
\qquad
m \in \{S,R\},
\end{equation}

where $l$ indexes transformer layers and $g$ indexes KV groups. When tokenizers differ, we compare positions ending at shared character boundaries in the source text (Appendix~\ref{app:extensions}). The adapter produces $\widetilde{\mathcal{C}}_R(\bm X)=\mathcal T_{\bm \theta}(\mathcal{C}_S(\bm X))$ and is optimized to reconstruct the receiver's own cache:
\begin{equation}
\mathcal{L}_{\mathrm{rec}}
=
\sum_{l,g}
\left(
\left\|
\widetilde{\bm K}_R^{(l,g)}
-
\bm K_R^{(l,g)}
\right\|_2^2
+
\left\|
\widetilde{\bm V}_R^{(l,g)}
-
\bm V_R^{(l,g)}
\right\|_2^2
\right).
\end{equation}

The reconstruction target encourages the adapter to retain information the receiver would encode from $\bm X$. This is motivated by the context-unaware setting, where the receiver must use the transformed cache in place of direct access to that input.

\paragraph{Phase II: generation-oriented communication.}

After reconstruction pretraining, we optimize the transformed cache for downstream generation under both communication regimes. For each training example, the receiver-side context $\bm X_R$ is drawn from a mixture of context-aware and context-unaware prompts,
\begin{equation}
\bm X_R
=
\begin{cases}
\bm X, & \text{context-aware},\\
\emptyset, & \text{context-unaware},
\end{cases}
\end{equation}

and the same adapter $\mathcal T_{\bm \theta}$ is updated across both cases:
\begin{equation}
\mathcal{L}_{\mathrm{gen}}
=
-
\sum_t
\log
p_{\mathcal{A}_R}
\left(
y_t
\mid
y_{<t},
\widetilde{\mathcal{C}}_R(\bm X),
\bm X_R
\right).
\end{equation}

The mixture exposes the same adapter to both uses of the cache: supplementing an observed input and providing source information that is otherwise unavailable. Reconstruction encourages agreement with receiver-native states, while generation training optimizes their usefulness for prediction.

\paragraph{Multiple independently encoded memories.}
For the three-agent task, two senders encode their documents separately without seeing the later question. Since they use the same sender model, the same adapter transforms both caches. We remove sender key rotations, transform each content cache, concatenate along the sequence dimension, and apply receiver key rotations at consecutive positions over the combined memory. Values are concatenated without rotary encoding. Phase II trains this two-memory interface conditioned on the receiver's question and local documents, using task-specific supervision rather than the original aware/unaware mixture.

\subsection{Cache Transformation Architecture}

Sender and receiver caches may differ in layer count, head dimension, KV-group organization, and positional encoding. Matching tensor shapes alone is insufficient because the models also learn different representations. The adapter handles these differences through positional disentanglement, fixed layer and group routing, and learned transformations.

\paragraph{Position-disentangled cache transformation.}
Rotary positional embeddings entangle keys with model-specific phase rotations. Removal and restoration below apply to keys; values are transformed without rotary encoding. Since dense communication should align the information stored in the cache rather than copy the sender's positional convention, $\mathcal T_{\bm \theta}$ first maps caches into a position-disentangled space:
\begin{equation}
\widehat{\mathcal{C}}_m
=
\mathrm{RemoveRoPE}_m
\left(
\mathcal{C}_m
\right),
\qquad
m \in \{S,R\}.
\end{equation}

The cross-model transformation is applied to $\widehat{\mathcal{C}}_S$ to produce a receiver-side content cache $\mathcal{C}_R'$. The final communicated cache restores the receiver's positional convention:
\begin{equation}
\widetilde{\mathcal{C}}_R
=
\mathrm{AddRoPE}_R
\left(
\mathcal{C}_R'
\right).
\end{equation}

The adapter thus transforms content without sender key rotations and places the result at the receiver's memory positions.

\paragraph{Layer alignment across different depths.}
Let $L_S$ and $L_R$ denote the sender and receiver depths. For each receiver layer $l$, we pair it with a sender layer through a monotonic depth-preserving map
\begin{equation}
a(l)
=
\min
\left(
\mathrm{round}
\left(
\frac{lL_S}
     {L_R}
\right),
L_S-1
\right),
\qquad
l=0,\ldots,L_R-1.
\end{equation}

This fixed map pairs layers at similar relative depths.

\paragraph{KV-group transformation with structured gates.}
The adapter transforms KV heads; query heads are not stored in the cache. For receiver group $g$ in layer $l$, it routes one sender group with keys and values $\widehat{\bm K}_S^{(a(l),\pi_l(g))},\widehat{\bm V}_S^{(a(l),\pi_l(g))}\in\mathbb R^{B\times S\times d_S}$, where $B$ is batch size and $S$ is sequence length. Each token vector is mapped to $\mathbb R^{d_R}$ and placed in receiver slot $(l,g)$, which the receiver's native query heads attend to.

Let $G_S$ and $G_R$ denote the number of sender and receiver KV groups per layer. For receiver group $g \in \{0,\ldots,G_R-1\}$, the fixed head routing is
\begin{equation}
\pi_l(g)
=
g \bmod G_S,
\end{equation}

so $\pi_l(g)$ maps a receiver KV group index to a sender KV group index in the aligned sender layer $a(l)$. If the layouts match, $\pi_l(g)=g$. If the receiver has more KV groups, sender groups are reused cyclically; if the sender has more KV groups, only the sender groups addressed by the receiver indices are injected. This routing is fixed, while the transformation parameters and gates below are learned. Each receiver KV group is produced from its routed sender-side content cache:
\begin{equation}
\left(
\bm K_R'^{(l,g)},
\bm V_R'^{(l,g)}
\right)
=
\gamma^{(l,g)}
\cdot
\left(
T_{K,\bm \theta}^{(l,g)}
\left(
\widehat{\bm K}_S^{(a(l),\pi_l(g))}
\right),
T_{V,\bm \theta}^{(l,g)}
\left(
\widehat{\bm V}_S^{(a(l),\pi_l(g))}
\right)
\right),
\end{equation}

where $\gamma^{(l,g)} \in [0,1]$ is a learnable gate for the receiver-side group. In this architecture, $T_{K,\bm \theta}^{(l,g)}$ and $T_{V,\bm \theta}^{(l,g)}$ are separate MLPs for each receiver KV group, applied to each token's key and value vectors:
\begin{equation}
T_{\star,\bm \theta}^{(l,g)}
(\bm z)
=
\bm W_{\star,2}^{(l,g)}
\,
\sigma
\left(
\bm W_{\star,1}^{(l,g)} \bm z
+
\bm b_{\star,1}^{(l,g)}
\right)
+
\bm b_{\star,2}^{(l,g)},
\end{equation}

Here $\star\in\{K,V\}$, $\sigma$ is GELU, and $\bm z\in\mathbb R^{d_S}$ is a sender key or value vector. The projections $\bm W_{\star,1}^{(l,g)}\in\mathbb R^{16d_S\times d_S}$ and $\bm W_{\star,2}^{(l,g)}\in\mathbb R^{d_R\times16d_S}$ expand the head dimension by $16$ and map to the receiver dimension. Parameters are shared across tokens but separate for each $(l,g,\star)$, including groups routed from the same sender group.

The retention study suggests that cache groups differ in their task relevance under the tested interventions. Separate transformations and gates let the adapter adjust each receiver group while reconstructing states across the cache.

\begin{table*}[t]
\centering
\resizebox{\textwidth}{!}{
\begin{tabular}{l l ccc ccc c}
\toprule
\multirow{2}{*}{\textbf{Pair}}
& \multirow{2}{*}{\textbf{Method}}
& \multicolumn{3}{c}{\textbf{In-domain}}
& \multicolumn{3}{c}{\textbf{OOD}}
& \multirow{2}{*}{\textbf{TFLOPs} $\downarrow$} \\
\cmidrule(lr){3-5} \cmidrule(lr){6-8}
&
& \textbf{GSM8K}
& \textbf{MATH-500}
& \textbf{ARC-C}
& \textbf{MMLU-Redux}
& \textbf{MedQA}
& \textbf{OpenBookQA}
& \\
\midrule
\multicolumn{9}{l}{\textit{Within-family: Qwen3 sender $\to$ Qwen3 receiver}} \\

\multirow{4}{*}{4B $\rightarrow$ 8B}
& Receiver-only & 81.10 & 49.20 & 91.00 & 72.10 & 53.00 & 91.20 & 19.85 \\
& T2T           & 88.10 & 76.00 & 91.74 & \textbf{80.75} & \textbf{67.40} & 90.40 & 37.73 \\
& C2C           & 77.86 & 44.20 & 86.09 & 75.87 & 56.87 & 85.60 & \textbf{4.50} \\
& Ours          & \textbf{92.95} & \textbf{82.00} & \textbf{93.69} & 78.52 & 67.24 & \textbf{91.20} & 12.56 \\
\midrule

\multirow{4}{*}{4B $\rightarrow$ 14B}
& Receiver-only & 83.70 & 46.40 & 92.60 & 71.80 & 64.70 & 91.80 & 31.16 \\
& T2T           & 92.34 & 77.80 & 92.00 & \textbf{82.87} & 71.17 & 92.00 & 56.24 \\
& C2C           & 82.34 & 44.20 & 92.43 & 72.76 & 63.00 & 87.00 & \textbf{6.64} \\
& Ours          & \textbf{93.86} & \textbf{86.00} & \textbf{94.20} & 78.57 & \textbf{71.96} & \textbf{93.60} & 21.54 \\
\midrule

\multirow{4}{*}{8B $\rightarrow$ 4B}
& Receiver-only & 82.40 & 44.20 & 89.20 & 65.00 & 47.70 & 88.00 & 9.18 \\
& T2T           & 89.39 & 63.20 & 90.96 & \textbf{79.39} & \textbf{66.46} & 87.00 & 33.91 \\
& C2C           & 72.48 & 37.40 & 86.78 & 70.19 & 55.07 & 77.20 & \textbf{3.43} \\
& Ours          & \textbf{91.81} & \textbf{83.40} & \textbf{93.00} & 77.86 & 66.30 & \textbf{89.60} & 7.95 \\
\midrule

\multirow{4}{*}{8B $\rightarrow$ 14B}
& Receiver-only & 83.70 & 46.40 & 92.60 & 71.80 & 64.70 & 91.80 & 31.16 \\
& T2T           & 93.93 & 75.00 & 91.74 & \textbf{84.13} & \textbf{72.66} & 92.80 & 67.08 \\
& C2C           & 82.26 & 43.20 & 92.43 & 73.81 & 64.65 & 87.00 & \textbf{7.42} \\
& Ours          & \textbf{94.09} & \textbf{85.00} & \textbf{94.37} & 77.38 & 70.46 & \textbf{93.40} & 21.79 \\
\midrule

\multirow{4}{*}{14B $\rightarrow$ 4B}
& Receiver-only & 82.40 & 44.20 & 89.20 & 65.00 & 47.70 & 88.00 & 9.18 \\
& T2T           & 90.60 & 60.20 & 89.83 & \textbf{80.61} & \textbf{69.91} & 85.00 & 43.48 \\
& C2C           & 70.58 & 36.60 & 85.83 & 69.73 & 53.42 & 76.40 & \textbf{5.01} \\
& Ours          & \textbf{91.13} & \textbf{82.60} & \textbf{91.89} & 77.66 & 63.00 & \textbf{88.80} & 10.18 \\
\midrule

\multirow{4}{*}{14B $\rightarrow$ 8B}
& Receiver-only & 81.10 & 49.20 & 91.00 & 72.10 & 53.00 & 91.20 & 19.85 \\
& T2T           & 91.58 & 73.00 & 92.35 & \textbf{83.43} & \textbf{72.90} & 90.20 & 55.50 \\
& C2C           & 76.35 & 43.00 & 89.39 & 75.99 & 62.69 & 85.60 & \textbf{7.83} \\
& Ours          & \textbf{92.95} & \textbf{81.40} & \textbf{93.77} & 78.04 & 70.38 & \textbf{92.60} & 15.64 \\

\midrule
\multicolumn{9}{l}{\textit{Cross-family: Mistral-7B-Instruct-v0.3 $\to$ Qwen3-4B}} \\
& Receiver-only & 82.40 & 44.20 & 89.20 & 65.00 & 47.70 & \textbf{88.00} & 9.18 \\
& T2T & 72.25 & 46.20 & 78.33 & 66.03 & 54.20 & 75.20 & 15.17 \\
& C2C & 51.78 & 26.60 & 83.11 & 66.55 & 58.21 & 73.20 & 14.17 \\
& Ours & \textbf{90.37} & \textbf{79.60} & \textbf{91.30} & \textbf{74.54} & \textbf{61.98} & 83.80 & \textbf{7.62} \\
\bottomrule
\end{tabular}

}
\small
\caption{\textbf{Context-aware accuracy (\%)} for six Qwen3 pairs and Mistral-to-Qwen transfer. Ours uses final Phase-II checkpoints. Estimated inference TFLOPs summarize model computation. The cross-family receiver-only row uses the same Qwen3-4B baseline as the within-family evaluation. Additional training protocols are detailed in Appendix~\ref{app:extensions}; compute accounting is given in Appendix~\ref{app:efficiency_setup}.}
\label{tab:main_context_multitask_extended}

\end{table*}

\section{Experiments}
\label{sec:experiments}
We first evaluate individual cache transfers with and without receiver access to the source input, including a cross-family pair. We then test whether a receiver can combine memories independently constructed by two agents before its question arrives.

\paragraph{Training Data.}
For the six-pair Qwen3 study, Phase I uses 30,000 reconstruction steps on OpenHermes. Phase II uses 2,000 steps on GSM8K~\citep{cobbe2021training}, MATH-algebra~\citep{hendrycks2021measuring}, and ARC-Challenge~\citep{clark2018think}, supervised by traces generated by the receiver with gold-answer guidance (Appendix~\ref{app:phase2_traces}). The three-agent and cross-family training protocols are detailed in Appendix~\ref{app:extensions}.

\paragraph{Experimental Setup.}
We evaluate the final Phase-II checkpoint for each trained adapter.
We evaluate our learned cache transformation across all six directed pairs among Qwen3-4B, 8B, and 14B~\citep{yang2025qwen3}, and additionally evaluate Mistral-7B-Instruct-v0.3 $\to$ Qwen3-4B. Evaluation is performed on three in-domain tasks (GSM8K, MATH-500, ARC-C) and three held-out multiple-choice benchmarks (MMLU-Redux~\citep{gema2025we}, MedQA~\citep{jin2021medqa}, OpenBookQA~\citep{OpenBookQA2018}). We compare three baselines against our method: \textbf{Receiver-only} (single-agent with the receiver model), \textbf{T2T} (text-based communication), and \textbf{C2C} \citep{fu2025cache} (learned cache projection and fusion). Both regimes from \Cref{sec:analysis} are evaluated: \emph{context-aware}, where the receiver also sees the question, and \emph{context-unaware}, where the receiver only sees the transferred signal. We estimate inference FLOPs from model sizes and token counts (Appendix~\ref{app:efficiency_setup}). Appendix~\ref{app:extensions} describes the three-agent and cross-family protocols.

\subsection{Context-aware Results}

\Cref{tab:main_context_multitask_extended} tests whether transferred representations help when the receiver already has the question. Our method improves over T2T on all three in-domain benchmarks across all six Qwen3 pairs, including gains of \(+6.00\) to \(+22.40\)\,pp on MATH-500. The mathematical reasoning gains suggest that dense alignment can strengthen reasoning-oriented communication even when the receiver already has the task input.

Our method also outperforms C2C on all six benchmarks across all six Qwen3 pairs. C2C modulates the receiver's cache through projected sender states, head-level weighting, and gated residual fusion~\citep{fu2025cache}. Our approach instead reconstructs receiver-compatible states across KV groups, then refines their transformations and weights through generation supervision. The gains show that the reconstruction-and-generation interface provides useful representations even when the receiver already observes the input. Generalization remains task-dependent: our method improves over T2T on OpenBookQA across all six pairs, while T2T leads on MMLU-Redux and five of six MedQA comparisons. Mistral-to-Qwen transfer exceeds T2T on all six benchmarks and the Qwen3-4B receiver-only baseline on five, demonstrating that the shared-input benefit extends to Mistral-to-Qwen transfer. Our method uses fewer inference FLOPs than T2T across all six Qwen3 pairs, avoiding sender decoding and receiver re-encoding of generated messages. C2C uses fewer inference FLOPs, but with much lower accuracy.

\begin{table*}[t]
\centering

\renewcommand{\arraystretch}{0.95}

\resizebox{\textwidth}{!}{
\begin{tabular}{l l ccc ccc c}
\toprule
\multirow{2}{*}{\textbf{Pair}}
& \multirow{2}{*}{\textbf{Method}}
& \multicolumn{3}{c}{\textbf{In-domain}}
& \multicolumn{3}{c}{\textbf{OOD}}
& \multirow{2}{*}{\textbf{TFLOPs} $\downarrow$} \\
\cmidrule(lr){3-5} \cmidrule(lr){6-8}
&
& \textbf{GSM8K}
& \textbf{MATH-500}
& \textbf{ARC-C}
& \textbf{MMLU-Redux}
& \textbf{MedQA}
& \textbf{OpenBookQA}
& \\
\midrule
\multicolumn{9}{l}{\textit{Within-family: Qwen3 sender $\to$ Qwen3 receiver}} \\

\multirow{3}{*}{4B $\rightarrow$ 8B}
& T2T-context-unaware  & 51.63 & 74.40 & 19.48 & 21.00 & 18.85 & 24.00 & 37.86 \\
& C2C-context-unaware  & 1.90 & 3.00 & 22.52 & 21.91 & 8.96 & 27.00 & 14.94 \\
& Ours-context-unaware & \textbf{91.43} & \textbf{78.80} & \textbf{91.38} & \textbf{74.59} & \textbf{61.82} & \textbf{88.80} & \textbf{9.42} \\
\midrule

\multirow{3}{*}{4B $\rightarrow$ 14B}
& T2T-context-unaware  & 56.79 & \textbf{75.20} & 23.39 & 23.06 & 27.65 & 27.20 & 60.83 \\
& C2C-context-unaware  & 0.00 & 0.00 & 10.70 & 10.26 & 12.80 & 16.40 & 20.05 \\
& Ours-context-unaware & \textbf{82.26} & 70.60 & \textbf{86.86} & \textbf{62.86} & \textbf{53.26} & \textbf{82.40} & \textbf{17.15} \\
\midrule

\multirow{3}{*}{8B $\rightarrow$ 4B}
& T2T-context-unaware  & 27.98 & 69.40 & 23.74 & 25.18 & 22.94 & 23.20 & 32.75 \\
& C2C-context-unaware  & 0.38 & 2.40 & 10.35 & 8.20 & 7.23 & 10.00 & 14.76 \\
& Ours-context-unaware & \textbf{91.36} & \textbf{81.60} & \textbf{93.60} & \textbf{77.06} & \textbf{64.57} & \textbf{90.00} & \textbf{6.56} \\
\midrule

\multirow{3}{*}{8B $\rightarrow$ 14B}
& T2T-context-unaware  & 30.86 & \textbf{70.40} & 22.96 & 22.80 & 27.81 & 27.40 & 72.06 \\
& C2C-context-unaware  & 0.00 & 0.00 & 7.39 & 7.14 & 6.36 & 7.60 & \textbf{9.84} \\
& Ours-context-unaware & \textbf{81.58} & 65.00 & \textbf{88.48} & \textbf{54.92} & \textbf{57.74} & \textbf{83.80} & 15.85 \\
\midrule

\multirow{3}{*}{14B $\rightarrow$ 4B}
& T2T-context-unaware  & 18.57 & \textbf{66.60} & 20.61 & 24.52 & 21.13 & 20.20 & 41.95 \\
& C2C-context-unaware  & 0.68 & 0.40 & 4.96 & 6.11 & 0.63 & 6.60 & 17.93 \\
& Ours-context-unaware & \textbf{82.49} & 64.00 & \textbf{88.48} & \textbf{67.15} & \textbf{59.54} & \textbf{85.40} & \textbf{8.90} \\
\midrule

\multirow{3}{*}{14B $\rightarrow$ 8B}
& T2T-context-unaware  & 18.80 & 66.40 & 22.87 & 21.41 & 25.22 & 26.20 & 53.80 \\
& C2C-context-unaware  & 0.00 & 0.20 & 2.78 & 5.34 & 5.18 & 2.20 & 24.77 \\
& Ours-context-unaware & \textbf{84.15} & \textbf{68.40} & \textbf{88.99} & \textbf{67.84} & \textbf{54.20} & \textbf{88.00} & \textbf{13.28} \\

\midrule
\multicolumn{9}{l}{\textit{Cross-family: Mistral-7B-Instruct-v0.3 $\to$ Qwen3-4B}} \\
& T2T-context-unaware & 62.24 & 34.60 & 75.43 & \textbf{61.47} & \textbf{51.77} & \textbf{74.60} & 14.41 \\
& C2C-context-unaware &      2.73 & 3.00 & 31.91 & 27.89 & 25.61 & 32.20 & 13.88 \\
& Ours-context-unaware & \textbf{64.97} & \textbf{54.40} & \textbf{77.13} & 55.74 & 41.01 & 73.00 & \textbf{6.64} \\

\bottomrule
\end{tabular}
}
\small
\caption{\textbf{Context-unaware accuracy (\%)} with no receiver access to the source input. T2T sends text; C2C and Ours send KV caches. Cross-family T2T sends only the generated message. Checkpoint and compute conventions follow \Cref{tab:main_context_multitask_extended}; Appendix~\ref{app:extensions} describes the cross-family evaluation protocol.}
\label{tab:context_unaware_multitask_extended}
\end{table*}

\subsection{Context-unaware Results}
\label{sec:context-unaware-results}

\Cref{tab:context_unaware_multitask_extended} reports the context-unaware regime, where only the sender observes the question and the receiver must produce the answer from the communicated signal alone. Under the context-unaware condition, the baselines degrade substantially: T2T-context-unaware achieves \(18.57\)--\(56.79\%\) on GSM8K and remains substantially below our method across the multiple-choice benchmarks, while C2C-context-unaware achieves only \(0.00\)--\(3.00\%\) on MATH-500 and similarly degrades across the remaining tasks.

In contrast, our learned transformation remains effective when the receiver has no access to the original question. Across the six Qwen3 model pairs, Ours-context-unaware achieves \(81.58\)--\(91.43\%\) on GSM8K, \(64.00\)--\(81.60\%\) on MATH-500, \(86.86\)--\(93.60\%\) on ARC-C, \(54.92\)--\(77.06\%\) on MMLU-Redux, \(53.26\)--\(64.57\%\) on MedQA, and \(82.40\)--\(90.00\%\) on OpenBookQA. The aligned cache therefore retains substantial source-side information even when the receiver must generate without the original question. Inference FLOPs are also lower than in the corresponding context-aware configurations. Removing the question reduces receiver prefill, while differences in generated output length also affect the totals. For example, \(8\text{B}\rightarrow4\text{B}\) achieves \(91.36/81.60/93.60\%\) on GSM8K/MATH-500/ARC-C at \(6.56\) inference TFLOPs, compared with \(89.39/63.20/90.96\%\) for context-aware T2T at \(33.91\) inference TFLOPs. For Mistral-to-Qwen context-unaware transfer, Ours exceeds text communication on the three in-domain benchmarks but trails it on MMLU-Redux, MedQA, and OpenBookQA (\Cref{tab:context_unaware_multitask_extended}). Cross-family transfer thus preserves the in-domain advantage over text communication, while the held-out results indicate weaker generalization beyond the training tasks.

\subsection{Partial-Aware Results: Three-Agent Document Sharing}
\label{sec:three_agent}
A user may share useful information with several agents and later ask another agent a question that draws on those earlier interactions. The later agent must combine information held by the others, whose memories were formed before the question was known. We instantiate the partial-aware regime (\Cref{subsec:communication-regimes}) with HotpotQA~\citep{yang2018hotpotqa}: agents A and B each observe one supporting document and separate distractors, without seeing the question. Agent C later receives the question and remaining distractors, then answers using information communicated by both agents. C thus has local context but relies on A and B for the supporting evidence.

A and B use the same sender model, and C uses a different model as receiver; A/B $\to$ C denotes sender and receiver sizes. We evaluate three configurations at the final checkpoint after task-specific two-memory training. We compare \textbf{Text communication}, which forwards each sender's complete documents with source-agent labels; \textbf{Latent (ours)}, which transfers independently encoded, aligned KV caches; and \textbf{Receiver-only}, a single instance of C given the question and all raw documents together. Evaluation excludes the development pool, leaving 7,241 examples; Appendix~\ref{app:extensions} gives additional implementation details.

\begin{table}[t]
\centering\small
\begin{tabular}{lccc}
\toprule
A/B $\to$ C & Text communication & Latent (ours) & Receiver-only \\
\midrule
8B$\to$4B & 23.84 / 30.45 & \textbf{55.97} / \textbf{70.15} & 35.88 / 47.32 \\
4B$\to$8B & 47.53 / 61.53 & \textbf{57.84} / \textbf{71.86} & 53.49 / 68.10 \\
14B$\to$4B & 23.84 / 30.45 & \textbf{39.08} / \textbf{53.92} & 35.88 / 47.32 \\
\bottomrule
\end{tabular}
\caption{HotpotQA answer EM / F1 (\%). Text communication forwards each sender's complete documents with source-agent labels; Latent (ours) transfers aligned caches; Receiver-only receives all documents directly. Bold marks the best EM and F1 within each row.}
\label{tab:hotpot_three}

\end{table}

\Cref{tab:hotpot_three} shows that latent communication improves both exact match and F1 over Text communication and Receiver-only in all three configurations. For $8$B$\to4$B, it achieves $55.97\%$ exact match and $70.15\%$ F1, compared with $23.84/30.45\%$ for text communication and $35.88/47.32\%$ for Receiver-only. The receiver can therefore combine independently constructed memories to answer a question neither sender saw during encoding. Its advantage over a receiver given all documents also suggests that how evidence is represented and integrated matters beyond simply making it available.

\section{Conclusion}
\label{sec:conclusion}
We presented dense KV-cache alignment for sharing internal representations across language-model agents. The retention study motivates broad information preservation with group-specific transformations and learned gates, refined through generation supervision. Across six Qwen3 pairs, the method improves in-domain accuracy over text communication with fewer inference FLOPs in the context-aware setting, and outperforms both communication baselines in 33 of 36 context-unaware task--pair comparisons in our evaluation. In the partial-aware setting, combining independently encoded document memories improves HotpotQA exact match and F1 over text communication and Receiver-only in all three configurations. Mistral-to-Qwen results further demonstrate transfer across model families. Together, these findings support dense alignment as an interface for reasoning-oriented communication and for sharing information beyond an agent's local observations.

\paragraph{Limitations.}
The current interface is trained to align the internal representation spaces of heterogeneous agents, meaning that supporting a new model pair requires training a new adapter. Dense caches also incur a larger memory footprint than text, resulting in higher memory usage and communication overhead for longer contexts. In addition, our experiments are limited to topologies with up to three agents and a small number of cross-family settings. Evaluating broader communication topologies and more diverse cross-family configurations will be important future work toward establishing heterogeneous latent communication in more practical scenarios.

\paragraph{AI use statement.}
We used Codex and Claude Code to assist with code development, result checking, and manuscript editing. Codex also assisted with literature and reference checks, interpretation of results, and feedback on experimental methodology. We used Qwen3 models to generate synthetic, gold-answer-guided training traces for generation supervision, as described in Appendices~\ref{app:phase2_traces} and~\ref{app:extensions}. The authors take responsibility for the final manuscript, including all AI-assisted text, code, experimental results, and claims.

\printbibliography

\newpage
\appendix
\section{Extended Related Work}
\label{app:related_work}

\paragraph{Text-mediated multi-agent systems.}
Frameworks such as AutoGen~\citep{wu2024autogen} and MetaGPT~\citep{hong2024metagpt} coordinate specialized language-model agents through dialogue and workflows. Other work improves coordination, agent selection, and communication policies~\citep{chen2025optima,wang2025agentdropout,zhang2025cut,wan2025rema,zhang2025aflow,zhao2025sirius}. Text is interpretable and model-agnostic, but generated messages require decoding and receiver-side re-encoding. Generated summaries can omit source details; this limitation should be distinguished from forwarding the original evidence verbatim.

\paragraph{Choices of latent interface.}
Latent communication exchanges internal representations instead of generated natural-language messages~\citep{yu2026learning}. Interfaces include embedding-level signals~\citep{pham2024let}, hidden states or state trajectories~\citep{du2025enabling,ramesh2025communicating,tang2025augmenting,fein2025mixture,zou2026recursive}, and KV caches~\citep{shi2025kvcomm,fu2025cache,jin2026agent,li2026less,liu2024droidspeak,brito2026latent}. These interfaces can convey representations of observations and intermediate computations. KV caches retain contextual states that participate directly in attention, making them suitable for computation reuse. Selective transmission, compression, projection, and fusion offer different ways to adapt this interface~\citep{shi2025kvcomm,fu2025cache,li2026less}; their quality and computational tradeoffs depend on the receiver's available information and the evaluated protocol.

\paragraph{Cross-model alignment and receiver observations.}
Partial observability is a longstanding concern in multi-agent learning~\citep{omidshafiei2017deep}. Copies of the same model share a representation space, but matching architecture alone does not ensure alignment between separately trained models. Cross-model latent interfaces must account for differences in both tensor structure and learned coordinates~\citep{ramesh2025communicating,fu2025cache,liu2024droidspeak}.
Cache-to-Cache uses learned projections and gating to fuse caches under shared input~\citep{fu2025cache}, while concurrent work studies cross-model KV-cache transfer primarily for context or prefill reuse across model sizes, architectures, and families~\citep{heo2026cross,li2026universal,qu2026cachebridge}. Our focus is direct cache alignment that also supports generation without receiver access to the source input. This observation distinction is separate from model heterogeneity: a homogeneous or heterogeneous pair may operate in any of the regimes in \Cref{subsec:communication-regimes}. The controlled retention analysis isolates observation availability using self-communication, while the learned adapter addresses transfer across different models.

\section{Transformer Attention and KV-Cache Preliminaries}
\label{app:attention_preliminaries}

\paragraph{Queries, keys, and values.}
For an input of $N$ tokens, let $\bm H^{(l-1)}\in\mathbb R^{N\times d_{\mathrm{model}}}$ denote the hidden states entering attention at layer $l$. With grouped-query attention (GQA), there are $H^q$ query heads and $G$ KV groups. Each query head $h$ is assigned to a KV group $g(h)$. Before positional encoding, the projections are
\begin{align}
\bm Q^{(l,h)} &= \bm H^{(l-1)}\bm W_Q^{(l,h)},\\
\bm K^{(l,g)} &= \bm H^{(l-1)}\bm W_K^{(l,g)},\qquad
\bm V^{(l,g)} = \bm H^{(l-1)}\bm W_V^{(l,g)},
\end{align}
where each projection matrix has shape $d_{\mathrm{model}}\times d_{\mathrm{head}}$, and each projected tensor has shape $N\times d_{\mathrm{head}}$. Standard multi-head attention is the special case $G=H^q$; under GQA, multiple query heads share the same key and value tensors. After applying the model's positional operations to queries and keys, attention for query head $h$ is
\begin{equation}
\mathrm{Attention}^{(l,h)}
=\mathrm{softmax}\!\left(
\frac{\bm Q^{(l,h)}(\bm K^{(l,g(h))})^\top}{\sqrt{d_{\mathrm{head}}}}
+\bm M
\right)\bm V^{(l,g(h))},
\end{equation}
where $\bm M$ is the causal mask, with zero entries for allowed positions and $-\infty$ for masked positions. The notation in this equation denotes the position-encoded queries and keys; values do not receive RoPE in the models considered here.

\paragraph{Prefill and autoregressive caching.}
During prefill, the model processes the input and stores the key and value tensors at every layer. During subsequent decoding, it appends the new token's keys and values and reuses the stored tensors instead of recomputing projections for previous tokens. Attention still accesses the retained history; caching does not eliminate that attention cost. For model $m\in\{S,R\}$, the cache is
\begin{equation}
\mathcal C_m(\bm X)
=\{(\bm K_m^{(l,g)}(\bm X),\bm V_m^{(l,g)}(\bm X))\}_{l,g}.
\end{equation}
Including the batch axis, each layer's key or value cache has shape $[B,G,N,d_{\mathrm{head}}]$. Query-head fanout is not an additional cache axis. These dimensions and the learned representation geometry may differ between sender and receiver, motivating the transformation in \Cref{sec:method}.

\section{Phase-II Trace Construction}
\label{app:phase2_traces}

For the six-pair Qwen3 study, Phase II trains the cache transformation with a supervised generation loss, so each source context must be paired with a receiver-side target output. We construct these targets as \emph{receiver-self guided traces}: for each sender--receiver pair $\mathcal{A}_S\!\rightarrow\!\mathcal{A}_R$, the trace generator is the receiver model $\mathcal{A}_R$ itself. This choice avoids teaching the receiver to imitate a different model's wording or reasoning style after cache transfer; the adapter instead learns to produce receiver-native latent states that decode into outputs the receiver is already well suited to generate.

\paragraph{Trace-generation tasks.}
The Phase-II trace pool contains three in-domain training tasks: GSM8K, MATH-algebra, and ARC-Challenge. These are the same tasks used for multitask Phase-II training in the main experiments. MATH-500, MMLU-Redux, MedQA, and OpenBookQA are not used for Phase-II trace generation; they are held for evaluation, with MATH-500 replacing MATH-algebra as the reported mathematical reasoning benchmark.

\paragraph{Guided trace generation.}
For each training example $(\bm X,\bm y)$, we prompt the receiver model with the source question and the ground-truth answer, then ask it to produce a step-by-step solution that arrives at that answer. The resulting JSONL record contains the question, the gold answer, the generated solution trace, and a trace-mode tag. For multiple-choice tasks such as ARC-Challenge, the trace is relabeled to end with the canonical answer format \texttt{Answer: X}, matching the receiver-side evaluation prompt. This relabeling prevents the training target from using a format that the evaluator later rejects or scores inconsistently.

\paragraph{Pair-specific receiver traces.}
Because the trace generator is the receiver, different cross-model directions use different trace files. For example, $14$B$\rightarrow4$B uses Qwen3-4B-self traces, while $4$B$\rightarrow14$B uses Qwen3-14B-self traces. These traces are generated on the training split with guided decoding and then matched back to training examples by question text before Phase-II optimization.

\paragraph{Mixed-regime receiver prompts.}
During Phase II, we train the same adapter on both context-aware and context-unaware receiver prompts. We set the receiver prompt to be context-aware with probability $0.5$ and context-unaware with probability $0.5$. Equivalently, the receiver-side context variable in \Cref{sec:method} is sampled as $\bm X_R=\bm X$ half of the time and $\bm X_R=\emptyset$ half of the time. The target trace is unchanged across these two cases; only the receiver's direct access to the source context changes. This trains the cache to supplement an observed input or provide source information when that input is unavailable.

In the six-pair Qwen3 study, Phase II starts from the Phase-I reconstruction checkpoint and runs for $2000$ optimization steps with cross-entropy on the receiver-self trace tokens.

Gold answers guide training traces only and are not supplied at inference. The three-agent and cross-family data, observation splits, and training budgets are documented separately in Appendix~\ref{app:extensions}.

\section{Estimated Inference Compute and Token Accounting}
\label{app:efficiency}

We estimate inference compute for the six-pair Qwen3 study and break the FLOPs column of Tables~\ref{tab:main_context_multitask_extended}~and~\ref{tab:context_unaware_multitask_extended} into its four constituent per-side token counts: sender prefill $\ell^S_{\mathrm{pre}}$, sender decode $\ell^S_{\mathrm{dec}}$, receiver prefill $\ell^R_{\mathrm{pre}}$, and receiver decode $\ell^R_{\mathrm{dec}}$.

\subsection{Estimation recipe}
\label{app:efficiency_setup}

\paragraph{Sample composition.}
For each (pair, method, mode) cell we sample 17 examples from each of
six benchmarks (GSM8K, MATH-500, ARC-Challenge, MMLU-Redux, OpenBookQA,
MedQA) using a fixed seed, pooled to $\approx$102 samples per cell.
We report mean token counts in the table.

\paragraph{Inference recipe.}
We record sender and receiver prefill and decode token counts during single-example inference and use these counts to estimate inference compute. Adapter training, trace generation, and communication transport costs are outside the estimate.

\paragraph{FLOPs estimate.}
We use the standard ``2 parameters per token'' rule of thumb for the
forward pass:
\[
\widehat{\mathrm{FLOPs}} \;\approx\; 2\, N_S \, (\ell^S_{\mathrm{pre}} + \ell^S_{\mathrm{dec}})
\;+\; 2\, N_R \, (\ell^R_{\mathrm{pre}} + \ell^R_{\mathrm{dec}}),
\]
where $N_S,N_R$ are nominal model-size proxies ($4$, $8$, or $14$ billion parameters), as used by the summarizer. The estimator includes only these sender and receiver parameter--token terms. It does not explicitly count adapter operations, attention over the retained history, serialization, or network transfer. These omissions limit the estimate to model computation; end-to-end latency also depends on cache processing and transport.

\paragraph{Token-count instrumentation.}
Token counts are exact, not estimated from rendered text. We record
generated token IDs directly and compute per-sample prefill/decode
lengths from model outputs.

\paragraph{Bandwidth disclosure.}
The saved payload estimates range from approximately $27.2$ to $30.2$ MB per example for Ours and $3.17$ to $3.95$ kB for generated text across the original pair/regime aggregates (decimal units). These are estimated message sizes, not measured network traffic. Transport and dense-cache storage can affect deployment cost and latency even when the parameter--token FLOPs estimate is lower.

\subsection{Per-side breakdown}
\label{app:efficiency_breakdown}

Table~\ref{tab:efficiency_breakdown} reports the four per-side token counts plus the unweighted total \texttt{All tok} for every (pair, method, mode) cell. The structural mechanism behind each method's compute profile is then visible. Both Ours and C2C eliminate the sender's autoregressive reasoning ($\ell^S_{\mathrm{dec}}=0$); the receiver prefill is small for both ($\ell^R_{\mathrm{pre}} = 220$ for Ours, $117$ for C2C in w/ctx mode) -- substantially below T2T's $\ell^R_{\mathrm{pre}} \sim 1{,}100$--$1{,}300$ tokens of re-encoded sender text. The largest token-count difference between Ours and C2C is in the \emph{receiver decode} length: Ours produces explicit step-by-step reasoning before the answer ($\ell^R_{\mathrm{dec}} \approx 400$--$500$ in w/ctx mode), whereas C2C emits a more terse answer ($\ell^R_{\mathrm{dec}} \approx 80$--$170$). The lower compute estimate accompanies lower accuracy on most benchmarks (see Tables~\ref{tab:main_context_multitask_extended}~and~\ref{tab:context_unaware_multitask_extended}).

\begin{table*}[t]
\centering
\renewcommand{\arraystretch}{0.95}

\resizebox{.8\textwidth}{!}{
\begin{tabular}{l l l rrrr rr}
\toprule
\textbf{Pair} & \textbf{Method} & \textbf{Mode}
& \textbf{Spre} & \textbf{Sdec} & \textbf{Rpre} & \textbf{Rdec}
& \textbf{All tok} & \textbf{TFLOPs} \\
\midrule

\multirow{7}{*}{4B $\rightarrow$ 8B}
& Receiver-only         & w/ctx &   0 &     0 &   154 & 1{,}087 & 1{,}241 &           19.85 \\
& T2T                   & w/ctx & 180 & 1{,}033 & 1{,}255 &    497 & 2{,}965 &           37.73 \\
& T2T-context-unaware             & context-unaware & 180 & 1{,}077 & 1{,}180 &    558 & 2{,}995 &           37.86 \\
& C2C                   & w/ctx & 117 &     0 &   117 &    106 &    340 & \textbf{ 4.50} \\
& C2C-context-unaware             & context-unaware & 117 &     0 &    14 &    861 &    993 &           14.94 \\
& Ours                  & w/ctx & 184 &     0 &   220 &    473 &    877 &           12.56 \\
& Ours-context-unaware            & context-unaware & 184 &     0 &    63 &    433 &    681 &            9.42 \\
\midrule

\multirow{7}{*}{4B $\rightarrow$ 14B}
& Receiver-only         & w/ctx &   0 &     0 &   154 &    959 & 1{,}113 &           31.16 \\
& T2T                   & w/ctx & 180 & 1{,}020 & 1{,}242 &    423 & 2{,}866 &           56.24 \\
& T2T-context-unaware             & context-unaware & 180 & 1{,}052 & 1{,}155 &    666 & 3{,}053 &           60.83 \\
& C2C                   & w/ctx & 117 &     0 &   117 &     87 &    321 & \textbf{ 6.64} \\
& C2C-context-unaware             & context-unaware & 117 &     0 &    14 &    668 &    800 &           20.05 \\
& Ours                  & w/ctx & 184 &     0 &   220 &    497 &    901 &           21.54 \\
& Ours-context-unaware            & context-unaware & 184 &     0 &    63 &    497 &    744 &           17.15 \\
\midrule

\multirow{7}{*}{8B $\rightarrow$ 4B}
& Receiver-only         & w/ctx &   0 &     0 &   154 &    994 & 1{,}147 &            9.18 \\
& T2T                   & w/ctx & 180 & 1{,}074 & 1{,}296 &    433 & 2{,}984 &           33.91 \\
& T2T-context-unaware             & context-unaware & 180 & 1{,}037 & 1{,}139 &    520 & 2{,}877 &           32.75 \\
& C2C                   & w/ctx & 117 &     0 &   117 &     77 &    311 & \textbf{ 3.43} \\
& C2C-context-unaware             & context-unaware & 117 &     0 &    14 & 1{,}597 & 1{,}728 &           14.76 \\
& Ours                  & w/ctx & 184 &     0 &   220 &    405 &    810 &            7.95 \\
& Ours-context-unaware            & context-unaware & 184 &     0 &    63 &    388 &    636 &            6.56 \\
\midrule

\multirow{7}{*}{8B $\rightarrow$ 14B}
& Receiver-only         & w/ctx &   0 &     0 &   154 &    959 & 1{,}113 &           31.16 \\
& T2T                   & w/ctx & 180 & 1{,}066 & 1{,}288 &    396 & 2{,}930 &           67.08 \\
& T2T-context-unaware             & context-unaware & 180 & 1{,}073 & 1{,}175 &    682 & 3{,}111 &           72.06 \\
& C2C                   & w/ctx & 117 &     0 &   117 &     81 &    315 & \textbf{ 7.42} \\
& C2C-context-unaware             & context-unaware & 117 &     0 &    14 &    270 &    402 &            9.84 \\
& Ours                  & w/ctx & 184 &     0 &   220 &    453 &    857 &           21.79 \\
& Ours-context-unaware            & context-unaware & 184 &     0 &    63 &    398 &    645 &           15.85 \\
\midrule

\multirow{7}{*}{14B $\rightarrow$ 4B}
& Receiver-only         & w/ctx &   0 &     0 &   154 &    994 & 1{,}147 &            9.18 \\
& T2T                   & w/ctx & 180 &   923 & 1{,}145 &    430 & 2{,}677 &           43.48 \\
& T2T-context-unaware             & context-unaware & 180 &   886 &   988 &    525 & 2{,}579 &           41.95 \\
& C2C                   & w/ctx & 117 &     0 &   117 &     99 &    333 & \textbf{ 5.01} \\
& C2C-context-unaware             & context-unaware & 117 &     0 &    14 & 1{,}817 & 1{,}948 &           17.93 \\
& Ours                  & w/ctx & 184 &     0 &   220 &    407 &    811 &           10.18 \\
& Ours-context-unaware            & context-unaware & 184 &     0 &    63 &    405 &    652 &            8.90 \\
\midrule

\multirow{7}{*}{14B $\rightarrow$ 8B}
& Receiver-only         & w/ctx &   0 &     0 &   154 & 1{,}087 & 1{,}241 &           19.85 \\
& T2T                   & w/ctx & 180 &   890 & 1{,}112 &    483 & 2{,}666 &           55.50 \\
& T2T-context-unaware             & context-unaware & 180 &   887 &   989 &    505 & 2{,}562 &           53.80 \\
& C2C                   & w/ctx & 117 &     0 &   117 &    167 &    402 & \textbf{ 7.83} \\
& C2C-context-unaware             & context-unaware & 117 &     0 &    14 & 1{,}329 & 1{,}461 &           24.77 \\
& Ours                  & w/ctx & 184 &     0 &   220 &    435 &    839 &           15.64 \\
& Ours-context-unaware            & context-unaware & 184 &     0 &    63 &    444 &    692 &           13.28 \\

\bottomrule
\end{tabular}
}
\small
\caption{
\textbf{Per-side efficiency breakdown.} Mean per-sample token counts
on the sender (Spre / Sdec = prefill / decode) and receiver (Rpre /
Rdec) sides, pooled across the six benchmarks of
Tables~\ref{tab:main_context_multitask_extended}~and~\ref{tab:context_unaware_multitask_extended}.
\texttt{All tok} = $\ell^S_{\mathrm{pre}} + \ell^S_{\mathrm{dec}} + \ell^R_{\mathrm{pre}} + \ell^R_{\mathrm{dec}}$
(every token that touches either model).  TFLOPs computed via the
formula in Appendix~\ref{app:efficiency_setup}.  Bold marks the
lowest TFLOPs cell within each pair (across all methods + modes).
}
\label{tab:efficiency_breakdown}
\end{table*}

\subsection{Structural observations}
\label{app:efficiency_observations}

Three observations follow from Table~\ref{tab:efficiency_breakdown}.

\paragraph{(i) Both Ours and C2C eliminate sender decoding.} T2T averages approximately 900--1{,}080 sender-decode tokens under a 2,048-token generation budget. Ours instead encodes the question into a sender KV cache without autoregressive sender decoding ($\ell^S_{\mathrm{dec}}=0$); C2C similarly transfers a cache without sender generation.

\paragraph{(ii) Receiver prefill is small for both cache-based methods, $\sim$6$\times$ smaller than T2T.} Ours has $\ell^R_{\mathrm{pre}}=220$ (w/ctx) / $63$ (context-unaware); C2C has $117$ / $14$. Both stand in sharp contrast to T2T's $\ell^R_{\mathrm{pre}} \sim 1{,}100$--$1{,}300$ tokens of re-encoded sender message: the receiver no longer needs to ingest the sender's natural-language reasoning through its embedding pipeline.

\paragraph{(iii) Receiver decoding accounts for much of the compute difference between Ours and C2C.} In w/ctx mode Ours produces explicit step-by-step reasoning before the answer ($\ell^R_{\mathrm{dec}} \sim 400$--$500$); C2C emits a more terse output ($\ell^R_{\mathrm{dec}} \sim 80$--$170$). The lower decode length translates to a $\sim$2--3$\times$ smaller TFLOPs total for C2C w/ctx -- but at the cost of substantially lower accuracy on most benchmarks (Tables~\ref{tab:main_context_multitask_extended}~and~\ref{tab:context_unaware_multitask_extended}). In context-unaware mode C2C's receiver decode varies widely across pairs ($\ell^R_{\mathrm{dec}} \in [270, 1{,}817]$); Ours keeps a stable decode profile across all six pairs ($\ell^R_{\mathrm{dec}} \approx 400$), and Ours-context-unaware has the lowest estimated compute among the context-unaware methods in 5 of 6 pairs.

\subsection{Compute tradeoffs}

\paragraph{Receiver-only in 14B $\rightarrow$ 4B.} For $14$B$\rightarrow4$B, one of two directions with a 4B receiver, the bare Receiver-only baseline (4B alone, $\sim$994 receiver-decode tokens) costs $9.18$ TFLOPs. Adding our 14B sender prefill ($\sim$184 tokens) makes Ours marginally more expensive at $10.18$ TFLOPs in w/ctx mode, although Ours-context-unaware drops back to $8.90$. In every other direction Ours is cheaper than Receiver-only.

\paragraph{Scope of the compute comparison.}
The Qwen3 accuracy tables and efficiency summaries use the final Phase-II checkpoints. The reported task tradeoffs also differ: T2T is stronger on MMLU-Redux and most MedQA comparisons despite its higher estimated compute. Accuracy is evaluated on the benchmark sets, while compute is estimated on the smaller pooled sample. These measurements characterize accuracy and model-compute tradeoffs; they do not measure latency or total deployment cost.

\section{Compressed-Sensing Analysis Across Regimes}
\label{app:cs_setup}

Section~\ref{sec:analysis} uses post-hoc compressed sensing (CS)~\citep{bair2026compressed} to rank sender query heads by their sensitivity to ablation, then evaluate accuracy as a function of retained KV groups. This section documents the setup and results behind \Cref{fig:sparse-dense-contrast}.

\subsection{Self-communication setup}

Sender and receiver are the same Qwen3-4B model (homogeneous, identity KV pass-through): the sender encodes the question into a KV cache with zero sender decoding, and the receiver consumes that cache directly. Qwen3-4B has $36$ layers $\times$ $32$ query heads $= 1152$ query heads, organized into $36 \times 8 = 288$ KV groups under GQA (each KV head shared across $4$ query heads). The ranking stage ablates sender query-head outputs before the attention output projection; the retention stage instead zeros non-selected groups in the transmitted KV cache. Receiver attention is unchanged. These are distinct interventions: the head ranking is used as a heuristic for cache selection.

\paragraph{From query-head importance to KV-group importance.}
The CS measurements (Stage~1 below) perturb individual \emph{query heads}, so the Lasso problem is naturally posed at the $1152$-query-head granularity. The transmitted KV cache itself, however, has only $288$ KV groups. To go from one to the other, we sum the $4$ query-head Lasso coefficients within each GQA group:
\[
\mathrm{score}_{\text{KV}}^{(l, h_{\mathrm{KV}})}
\;=\;
\sum_{h_Q \,\in\, \mathrm{group}(h_{\mathrm{KV}})} \hat{x}_{l, h_Q},
\]
which gives a per-(layer, KV-head) score. More-negative scores indicate greater estimated importance under the ablation-mask convention below; Stage~2 retains the $K$ groups with the smallest scores.

\subsection{Stage 1: CS head ranking}

We take $M = 200$ binary ablation masks $\Phi \in \{0,1\}^{M \times N}$ over the $N = 1152$ query heads, with each row ablating approximately $5\%$ of heads. Here $\Phi_{mi}=1$ means head $i$ is ablated in measurement $m$, and zero means it is left intact. Ablation zeros the selected query-head output before the sender attention output projection. For each mask $m$, we run the full sender$\to$receiver pipeline on $100$ evaluation samples and record the resulting metric $y_m$ (accuracy on GSM8K / MATH-algebra / ARC-Challenge), then center it by the unablated baseline, $\tilde{y}_m = y_m - y_{\mathrm{baseline}}$. We then solve a Lasso regression for the per-head importance vector $\mathbf{x} \in \mathbb{R}^N$:
\[
(\hat{\mathbf{x}},\hat b) \;=\; \arg\min_{\mathbf{x}\in\mathbb{R}^N,\,b\in\mathbb R}\;
\frac{1}{2M}\,\bigl\|\tilde{\mathbf{y}} - \Phi\,\mathbf{x} - b\mathbf 1\bigr\|_2^2
\;+\; \alpha\,\|\mathbf{x}\|_1,
\qquad \alpha = 10^{-4},
\]
where $b$ is a fitted intercept. The additive surrogate is a ranking heuristic, not an assumption that attention heads act independently. We rank heads by $-\hat{x}_i$ (most-negative $\hat{x}_i$ indicates highest importance, since masking it most degrades accuracy).

\subsection{Stage 2: KV-group retention sweeps}
We evaluate the ranking by retaining selected groups in the transmitted cache and zeroing the remaining key and value entries. Receiver attention runs unmodified. The context-aware sweep includes $K\in\{5,10,20,50,200\}$; \Cref{fig:sparse-dense-contrast} displays $K=10,20,50$ together with the full-cache result at $K=288$ and a receiver-only reference at $K=0$. The context-unaware curve displays $K\in\{0,10,20,50,100,150,200,250,288\}$. The ranking is shared across regimes, but the displayed retention grids differ.

The context-aware receiver sees the question as well as the filtered cache; the context-unaware receiver is prompted to answer using the cache without seeing the question. The $K=0$ context-aware points are separately recorded receiver-only references: GSM8K and ARC-Challenge use the original paper's Qwen3-4B baseline, while MATH-algebra uses the corresponding historical single-agent result. The receiver-only references and retention scores come from separate evaluation runs, so differences from $K=0$ do not provide a controlled estimate of communication benefit.

\subsection{Random-filter baseline}

For the random-filter markers in \Cref{fig:sparse-dense-contrast}, we replace the CS-Lasso ranking with a uniform random selection of $K$ KV groups, repeat with $3$ seeds, and report the mean. The comparison tests whether the fitted ranking improves on random selection at the same retention level.

\subsection{Ranking limitations}
The Lasso surrogate uses 200 measurements for 1,152 query-head coefficients. Many lower-ranked coefficients are zero or close to zero, so aggregation and tie-breaking can affect which KV groups are retained as $K$ increases. A sparse fit alone does not establish statistical reliability or a recoverable number of task-relevant heads. We therefore interpret the curves as sensitivity to this selection procedure, rather than evidence of a minimal sufficient cache or independent causal head contributions.

\subsection{Context-Aware and Context-Unaware Results}

\subsubsection{Context-aware regime}
With $K=10$ retained groups, accuracy is $0.883/0.873/0.699$ on GSM8K/ARC-Challenge/MATH-algebra, compared with full-cache values of $0.917/0.880/0.698$. Random selection at $K=10$ gives similar results ($0.884/0.863/0.698$). The ranking has a larger effect at $K=20$ and $50$: at $K=50$, CS selection gives $0.876/0.862/0.577$, versus $0.605/0.583/0.484$ for random selection. These results show that high task accuracy can survive removal of most transmitted groups when the receiver also observes the input.

\subsubsection{Context-unaware regime}
Without receiver access to the question, accuracy remains low through $K=150$ and rises at higher retention. At $K=250$, the recorded GSM8K/ARC-Challenge/MATH-algebra scores are $0.834/0.904/0.686$, compared with $0.880/0.901/0.694$ at $K=288$. Each regime has its own full-cache reference. The comparison indicates greater sensitivity to KV-group removal when the cache supplies the source input.

\subsubsection{Implication for alignment}
The receiver's observations change what the transferred cache must provide. This motivates reconstructing receiver-compatible states across the cache, alongside generation supervision that trains the receiver to use them. KV-group transformations, gates, and positional handling implement this interface. The retention study motivates the overall design; component-level contributions require separate ablations.

\section{Additional Evaluation Protocols}
\label{app:extensions}

\subsection{Three-agent document memories}
The HotpotQA distractor split assigns one supporting document and separate distractors to each of A and B; C receives the remaining distractors and, later, the question. Supporting annotations define the partition but are not shown in prompts. For Latent (ours), senders encode only their own documents, without the question, answer, or the other sender's evidence. Each cache is transformed independently with the shared pair-specific adapter; transformed content caches are concatenated before receiver positional encoding. This is a three-agent topology with two instances of the same sender model. The partition requires two supporting documents and at least three distractors, leaving 7,368 of 7,405 validation examples.

The adapters undergo 30,000-step OpenHermes-2.5 reconstruction with a 500,000-example source pool (effective batch 32, learning rate $10^{-3}$, 1,000 warmup steps). Phase II uses 5,000 two-memory steps (effective batch 16, learning rate $10^{-4}$), with 20,042 training examples retained from a 22,000-example candidate pool after evidence and heldout filtering; 128 training examples are held out. Training supervision is Qwen3-4B-generated, gold-guided, greedy text with a 256-token budget; it is shared across pairs, providing a common supervision source for all receiver sizes. Gold answers are used for training supervision only. The fixed step-5,000 checkpoint is evaluated with greedy decoding and a 512-token answer budget.

Text communication (\Cref{sec:three_agent}) forwards A's and B's complete observed documents, labeled by source agent; no sender generation, evidence selection, or summarization is performed. Latent (ours) transfers the two independently encoded, aligned caches. Receiver-only is a single receiver given the question and all raw documents together, including both supporting documents, without sender agents. Text communication and Receiver-only therefore contain the same evidence in different presentations and are shared among pairs with the same receiver size. We report answer exact match and token F1, with F1 computed from token-count overlap. These answer-only metrics use the repository's text normalization.


\subsection{Cross-family protocol}
The sender is Mistral-7B-Instruct-v0.3 and the receiver is Qwen3-4B. Phase I compares caches at token positions ending at shared character boundaries on identical text, excluding unmatched positions. Shared character boundaries provide reconstruction targets at matching text prefixes despite different tokenizations. The run uses 30,000 reconstruction steps from a 500,000-example OpenHermes pool, followed by 2,000 generation-training steps on GSM8K, MATH-algebra, and ARC-Challenge. Generation training uses 10,335 examples matched to supervision traces by question text. Evaluation uses the final Phase-II checkpoint.

\section{Qualitative Latent Space Visualization}
\label{app:pca}
\begin{figure}[h]
    \centering
    \includegraphics[width=.45\linewidth]{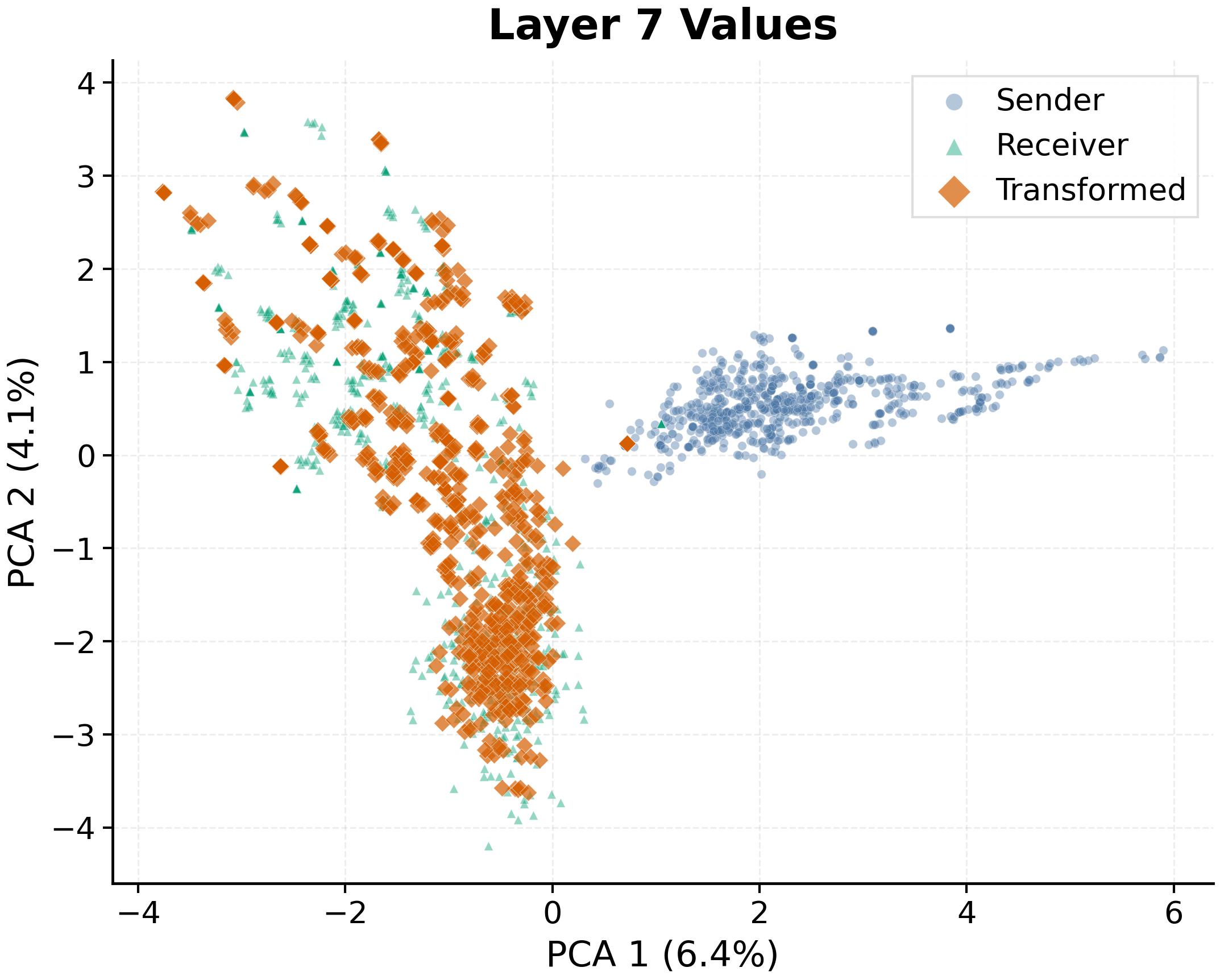}
    \includegraphics[width=.45\linewidth]{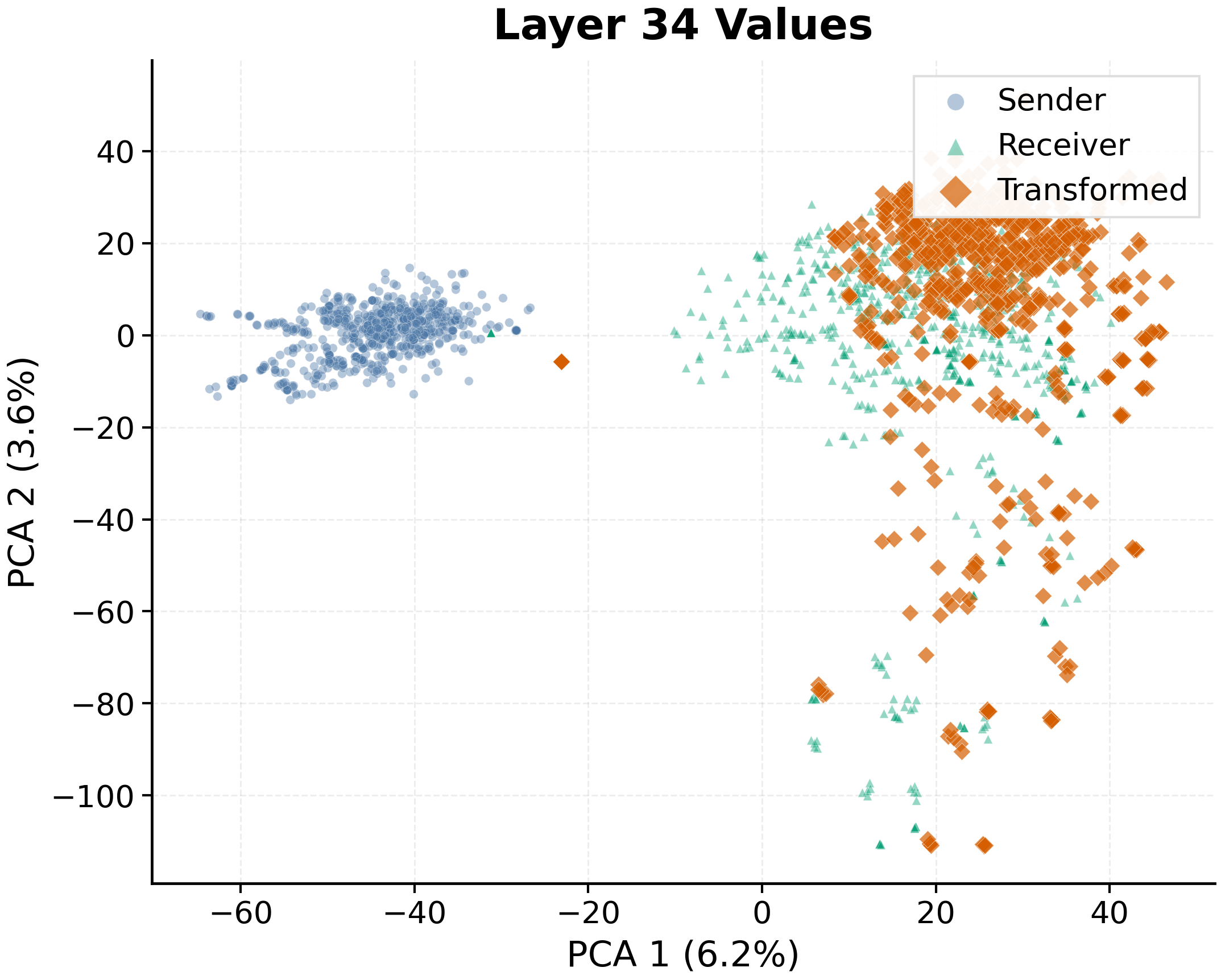}
    \caption{Two-dimensional PCA views of transformed sender and receiver-native value vectors for the same inputs at layers 7 and 34. The projections illustrate the geometry of aligned and receiver-native representations at two depths.}
    \label{fig:kv-cache-transform}
\end{figure}
\Cref{fig:kv-cache-transform} visualizes per-token value vectors in the receiver cache space. The projections suggest similarity between transformed and native representations at the displayed layers. As two-dimensional summaries, they complement the task-level evaluation but do not measure how much information the full cache preserves.

\end{document}